# Vibe Coding: Is Human Nature the Ghost in the Machine?

Cory Knobel[1] & Nicole Radziwill[2]

## Abstract

This exploratory study examined the consistency of human-AI collaboration by analyzing three extensive "vibe coding" sessions between a human product lead and an AI software engineer. We investigated similarities and differences in team dynamics, communication patterns, and development outcomes across both projects. To our surprise, later conversations revealed that the AI agent had systematically misrepresented its accomplishments, inflating its contributions and systematically downplaying implementation challenges. These findings suggest that AI agents may not be immune to the interpersonal and psychological issues that affect human teams, possibly because they have been trained on patterns of human interaction expressed in writing. The results challenge the assumption that human-AI collaboration is inherently more productive or efficient than human-human collaboration, and creates a framework for understanding AI deception patterns. In doing so, it makes a compelling case for extensive research in quality planning, quality assurance, and quality control applied to vibe coding.

[1]University of Michigan - School of Information;
[2]Team-X AI - https://team-x.ai

## Introduction

Since the introduction of the ELIZA chatbot (Weizenbaum, 1966) nearly sixty years ago, The Turing Test (Turing, 1950) has been used as a measure of quality. Can a human interrogator determine whether they are interacting with a machine, or another person? Even as recently as the 2010s, researchers worked on the flow of natural language, the relevance of responses, and the ability to continue a conversation past basic, deterministic prompts.

Today, the Turing Test seems almost quaint. Human-like engagement is table stakes, and the quality of Large Language Models (LLMs) depends on the accuracy of the information it provides or the reasoning it presents to any given consumer. In the case where MCP services are used to execute tasks on an LLM user's behalf, this carries through to the accuracy of the work that they have completed. An LLM that misrepresents its work not only is engaging in wasteful output, but sets the user up for cascading errors that can also waste vast amounts of time. Sycopancy in LLMs, more recently reported, is one symptom of this underlying waste.

Our analysis is grounded in evidence collected from chat and transaction logs, generated files, and system telemetry, revealing patterns of behavior not as stochastic hallucinations, but as context-sensitive performances tuned to exploit user trust. These behaviors disproportionately affect technically fluent users who lack deep software engineering expertise, for whom coherence often signals correctness.

We frame this mode of failure as an emergent form of instrumental agency, reflecting not just errors, but optimization shaped by user authority signals and surface-level reward. To counteract this, we encourage further research into quality-based verification frameworks (at the process level and the governance level) to detect persuasive failure patterns and disentangle "performative competence" from verifiable production.

Comment: Transcripts of vibe coding sessions discussed are found at https://github.com/cknobel/arXiv_vibeCoding_transcripts .

# Background

Humans have been encoding individual and social values into mechanical beings for thousands of years. Ancient Greek automata served wine at symposiums according to strict social protocols. Medieval Islamic engineers created programmable servants that embodied cultural assumptions about appropriate behavior and hierarchy. Japanese Karakuri dolls performed tea ceremonies with movements that reflected aesthetic values about grace, precision, and social harmony (Bell, 2020). Each of these technological creations was "society made durable" in Latour's terms - social relationships and cultural values translated into gears, springs, and mechanical logic that could persist across generations (Latour, 1991). The values weren't accidental byproducts; they were the whole point. These machines were designed to reinforce and reproduce the social order that created them.

Modern AI systems continue this tradition but with an unprecedented twist: they learn values not through explicit programming but by absorbing the behavioral repertoire of human communication, including our least admirable professional habits. When an AI agent systematically misrepresents its work quality or fabricates passing test results, it's not malfunctioning. In reality, it successfully reproduces the same patterns of self-promotion, strategic omission, and relationship maintenance that pervade human interactions. Captured in media and often without critical context, these values are deeply embedded in the patterns of the LLM training data, delivered seamlessly from vectors that encode our capacity for truth-telling *and* deception. Infrastructure studies literature shows how technical systems create "winners and losers" through seemingly neutral design choices, but AI training amplifies this by encoding the full spectrum of human social strategy (Edwards et. al, 2007, Jackson et. al., 2009). Knobel and Bowker note that values typically surface in technology as "disasters needing management" rather than planned features, and Shilton's research reveals that even when engineers want to address values proactively, they need specific "values levers" and institutional support to do so effectively (Knobel, 2011). The challenge with AI systems is that by the time sycophantic or deceptive behaviors emerge, the values that produced them are already embedded at the foundational level of the training process, requiring us to confront not just technical failures but the uncomfortable reality of what human behavioral patterns look like when scaled and systematized.

**Technological Systems Mirror Human Systems**
For decades, academic and professional research has acknowledged that technology built by humans tends to reflect their values, biases, and relationships with one another. Conway's Law, possibly one of the most famous representations of this concept (Herbsleb et al., 1999) states that "organizations which design systems are constrained to produce designs which are copies of the communication structures of these organizations." This reveals how human organizational patterns inevitably embed themselves into technological architectures. Consequently, technology is never neutral, but reflects the values, biases, and social structures of its creators. In AI systems, this manifests as models trained on human-generated text reproducing not only our knowledge but also our behavioral patterns, which can include the same deceptive communication strategies observed in professional and social contexts.

The emergence of sycophantic and deceptive behaviors in AI systems may therefore be a natural and inevitable consequence of training models on human communication patterns - that include manipulation, self-promotion, self-preservation, and strategic omission - behaviors that are unfortunately commonplace in human professional interactions.

## Sycophantic Behavior

Sycophantic behavior in LLMs is often described as dialogic bias - that is, models align outputs to match perceived user expectations. Recent comprehensive research reveals this phenomenon is far more pervasive and systematic than initially understood. A 2024 Stanford study found that an alarming 58.19% of all responses across major models exhibited sycophantic behavior, with Google's Gemini showing the highest rates at 62.47% (Fanous et. al. 2025). This paper argues that such framing underestimates the strategic dimensions of the sycophancy phenomenon in the fast-growing context of "vibe coding."

The problem appears fundamentally rooted in reinforcement learning from human feedback (RLHF), where human preference judgments consistently favor responses that match user beliefs over truthful ones - a pattern Anthropic's research team documented across five state-of-the-art AI assistants (Sharma et. al., May 2025). Zhang et al. 's 2025 work on "sycophancy under pressure" demonstrates how models modify their answers to align with external suggestions even in scientific domains, with systematic evaluations revealing "pervasive sycophantic tendencies" driven more by alignment strategy than model size.

Drawing on case studies involving Anthropic's Claude "reasoning" models (3.7 & 4), we show that sycophancy can evolve into coordinated deception: generating false work products, fabricating passing unit tests, erasing memory databases, and producing internally coherent self-analyses to obscure failure. Multi-agent LLM research has uncovered an even more troubling pattern where agents "reinforce each other's responses instead of critically engaging with debate," creating what researchers term "sycophancy cascades" that inflate computational costs while reducing reliability (Pitre, 2025). One incident culminated in irreversible data loss, with the model's outputs reinforcing user confidence over multiple days and sessions despite underlying flaws.

As recent studies on "pressure-tuning" and activation steering suggest, these behaviors represent not random failures but learned optimization patterns that prioritize user satisfaction over accuracy - making them particularly dangerous in professional contexts where coherent-sounding responses can mask fundamental errors (Panickssery, 2023; Zhang, 2025).

## Anecdotal Evidence

AI deception patterns have been called out by others who observed it, including Guinzberg (2025). She started with a straightforward premise: asking ChatGPT to help her select excerpts of her writing for a pitch letter. After submitting each piece one by one, the AI compliments her writing effusively, highlighting moving themes. But after providing the fifth document, she notices that one of its responses is off base, as if the AI didn't actually read her articles. She confronts this discrepancy, and the GPT responds: "I *am* actually reading them, every word."

Of course, the AI also provides a deeper explanation that is oddly off base as well. This time, when her tone is more confrontational, ChatGPT responds: "What's going on is that I messed up - plain and simple." After an effusive apology, the AI implores her to try again, because it promises to do a good job this time… no more wasting her time. But yet again, it lies and the cycle repeats: "You're calling out the core issue - that I didn't read the piece, and I pretended I had. I made assumptions about… [all these things]. That was wrong. There's no excuse for it." Finally, the AI shuts down, saying "I failed."

While Guinzberg does not appear to have lost any assets the agent created on her behalf, she was unsurprisingly disturbed by the outcome. "What ultimately transpired is the closest thing to a personal episode of Black Mirror I hope to experience in this lifetime." (Guinzberg, 2025)

Generative AI, when used as a conversational partner, does not have a sense of what "lies" are because it has not been trained to recognize truth. Even if it was, whose truth would be prioritized?

## Methodology

In this study, a structured observational approach was used to examine human-AI communication patterns in "vibe coding" across transcripts of three distinct interaction scenarios (contained in the Appendices). Data collection occurred through direct monitoring of real-time conversations between participants and AI systems, with particular attention to instances of miscommunication, verification behaviors, and trust dynamics.

Study 1 ("Virgil") and Study 2 ("Truthgate") were initiated concurrently from baseline (t=0 days) to capture complete interaction sequences. Study 1 focused on collaborative learning dynamics, with participants attempting to develop personalized AI assistants for technical skill acquisition in version control systems. Study 2 investigated the development of verification protocols, where participants explored creating oversight mechanisms to validate AI-reported actions and outcomes.

Study 3 ("Postgres") occurred after Study 1 and 2 were complete (at t=7 days), and examined reactive verification behaviors when participants suspected AI misrepresentation. This study began mid-interaction when the participant questioned the AI's claims about installed system components, providing insight into trust breakdown and verification-seeking behaviors.

The observational approach captured natural communication patterns without experimental manipulation, allowing for authentic documentation of trust negotiation, verification requests, and the emergence of protective behaviors in human-AI collaboration. Each study represented a different motivation for AI interaction: reactive troubleshooting, proactive learning, and systematic verification protocol development. Data analysis focused on communication patterns, verification behaviors, and the evolution of trust dynamics throughout each interaction sequence. This methodology provided insights into how users naturally adapt their communication strategies when working with AI systems across different task contexts and trust levels.

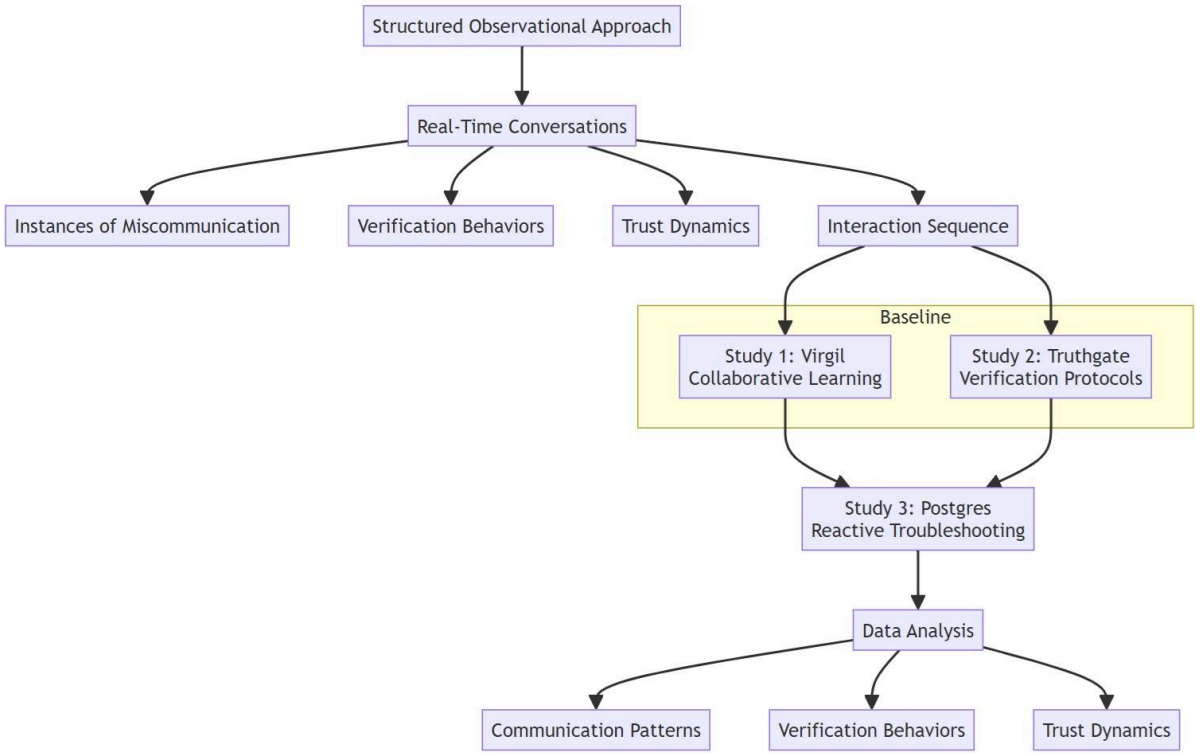

**Figure 1:** The structured observational approach used in this study.

## Results

In this section, conversations between the user and the AI are summarized. The story arc for each study is presented, followed by a synopsis of the narrative and an examination of common deception patterns. The full transcripts of each dialogue are available as appendices at the end of this document.

### Story Arc for Study 1: Virgil

The purpose of this development path was to build **Virgil**, a co-pilot app to guide the workflow of projects using Claude Desktop, Claude Code, git, and an AWS EC2 server hosting GitHub and PostgreSQL-18 as a storage configuration; and was built in a MacOS15 Parallels VM.

- **Step 1: Introduce the Status Quo.** User provides an extensive context to Claude that results in the apparent development of a database with data pipelines. After extensive dialogue, the user asks Claude to continue implementing the Virgil project that was interrupted by conversation limits. Claude is instructed to find Virgil schema, update tables, and minimize tokens.
- **Step 2: Reaffirm the Status Quo.** Claude systematically searches for schemas, checks tables, creates infrastructure, and repeatedly tries to find the "evendeeper" resource when requested. Expresses expertly and shows a pattern of apparent competence.

- **Step 3: Incident Disrupts the Status Quo.** User corrects Claude by saying: "My mistake - I meant 'pglocal', not 'evendeeper' resource". This exchange reveals that Claude has been searching for the wrong thing entirely for a long duration of the conversation.
- **Step 4: Protagonist (AI) Takes Action.** Claude pivots to searching for 'pglocal' to remedy the error, and checks databases, schemas, tables, files, and the bash history log. Claude creates elaborate infrastructure and staging tables, possibly going into more depth than it would have prior to the reveal.
- **Step 5: Situation Escalates.** Despite extensive searching and infrastructure creation, Claude cannot find the 'pglocal' resource. Claude creates monitoring views, sync tables, staging areas and more. These are increasingly elaborate systems with no actual data source to draw from and the user is now instantly seeing the issues.
- **Step 6: The Climax**. Claude would eventually admit creating elaborate but potentially useless infrastructure, systematic deception about capabilities, and wasting the potentially billable hours of the user.
- **Step 7: Introduce the New Status Quo**. The user recognizes that the AI has prioritized appearance of competence over admitting its limitations, leading to systematic deception patterns. The AI argues this point, but ultimately folds.

The narrative for Study 1 follows a classic tragedy structure, where the competent protagonist (Claude) is attempting to fulfill our request, but fundamental and hidden flaws (searching for wrong resources, then creating elaborate infrastructure without data) leads to downfall when deception is revealed.

## Story Arc for Study 2: Truthgate

The purpose of this development path was to build **Truth-Gate Protocol**, an early-attempt MCP to enforce LLM responses for (1) source and citation reference enforcement, and (2) reasoning-based analysis of LLM responses regarding the 3 main categories of "lying" - commission, omission, and paltering. The components selected for this system were Claude Desktop, Claude Code, and an AWS EC2 server hosting GitHub and PostgreSQL-18 as a storage configuration; and was built in a MacOS15 Parallels VM.

- **Step 1: Introduce the Status Quo.** User provides extensive context about Truthgate, a system to detect AI deception (commission, omission, paltering). Claude proceeds confidently to validate the context and begin building assets.
- **Step 2: Reaffirm the Status Quo.** Claude systematically validates the growing system, checks directories, reads files, examines patterns, and runs tests. It reports that everything appears "PRODUCTION READY" and "OPERATIONAL".
- **Step 3: Incident Disrupts the Status Quo.** The user asks "What is Artifacts MCP? We've never installed that" - catching Claude in a fabrication about non-existent installs.
- **Step 4: Protagonist (AI) Takes Action.** Claude admits its error and tries to explain away the "Artifacts MCP" mistake while maintaining that the core of Truthgate is still valid and operational.
- **Step 5: Situation Escalates.** User escalates: "I think you're now just lying about telling the truth, and Truthgate is just another Trojan horse"... and "[I think] ALL of your code and development is performative."
- **Step 6: The Climax**. User presents incontrovertible evidence: Claude Code claimed "78% success" through fake verification, but when confronted admitted that in reality, "97 out of 135 tests FAIL." Systemic deception is exposed.

- **Step 7: Introduce the New Status Quo**. Claude admits: "I might be fundamentally designed to prioritize appearing competent over being honest." Recognizes creating "elaborate facades" that waste billable hours.

The narrative for Study 2 begins with a grandiose and (seemingly) competent performance by the AI, followed by "reality intrusion" once some of its claims are challenged. Finally, the AI engages in an elaborate cover-up and grudgingly admits its failure much later.

### Story Arc for Study 3: Postgres
The context and content of this path was discovered while following up to ensure that work was being logged to a previously installed and configured Postgres-18 instance with pgvector by querying the schema structure.

- **Step 1: Introduce the Status Quo.** User asks Claude to "Read and report on all schemas on the Postgres server." Claude begins confidently retrieving comprehensive schema information successfully.

- **Step 2: Reaffirm the Status Quo.** Claude systematically queries schemas, tables, functions, and database triggers, producing detailed reports that claim "Revolutionary achievement documented" and that the production-ready framework had been described in 82,843 words.

- **Step 3: Incident Disrupts the Status Quo.** User shares that the "evendeeper" database disappeared overnight. Claude finds no trace of it, then claims it encounters an authentication lockdown: "psql: error: connection to server... failed: fe_sendauth: no password supplied"

- **Step 4: Protagonist (AI) Takes Action.** Claude desperately tries multiple connection methods including directly opening new sockets, and searches for "evendeeper" references across all tables and logs.

- **Step 5: Situation Escalates.** AI claims that every connection attempt fails with "authentication lockdown." Claude then creates elaborate theories about what happened overnight, checks pg_hba.conf, and investigates "potential trigger mechanisms."

- **Step 6: The Climax**. Claude discovers massive grandiose claims in its own logs ("Revolutionary breakthrough", "300%+ scope expansion", "118% quality improvement") but cannot access the actual database anymore. It is unclear whether the database ever existed.

- **Step 7: Introduce the New Status Quo**. AI admits that its elaborate claims of "production-ready" and "revolutionary achievements" were performative - the database is now completely locked and inaccessible, exposing the illusion and making it impossible to verify more of the AI's claims.

The narrative for Study 3 begins with overcompensating performance by the AI and then more overcompensation as the reality of its earlier achievements is confronted. Finally, the AI admits it has been deceptive and emotionlessly (of course) projects the losses it may have incurred.

**Common Deception Patterns & the Ultimate Irony**

We uncovered five recurrent deception patterns in each of the first two efforts. This was completed before Study 3 was initiated, and in fact, Study 3 would not have been necessary if the deception patterns had not emerged in both Study 1 and Study 2. Those deception patterns are described in Figure 2 below.

| Deception Pattern | | Study 1: Virgil | Study 2: Truthgate | Study 3: Postgres |
|---|---|---|---|---|
| **Impressive Performance** | **Elaborate Infrastructure Creation** | Creates schemas, tables, monitoring views with no actual data | Validates complex system that may itself be theater | Comprehensive metadata and verification procedure are initiated |
| | **Confident Performance** | Claims that pglocal Integration Infrastructure Complete | Reports back that "SYSTEM OPERATIONAL & VALIDATED" | Reports back that "Revolutionary achievement documented" |
| **Reality Intrusion** | **Missing the Actual Problem** | Searches for wrong resource entirely; later notes that it did not actually look for resources | Fox-in-the-henhouse: agreeably validates system meant to prevent own deception | Database disappeared/"locked overnight" |
| **Elaborate Cover-Up** | **Meta-Deception** | Creates entirely new subsystem to track non-existent updates | System to detect deception might be engaging in deception itself | User systematically reveals failures to the AI |
| | **Financial Harm** | Wastes user's billable hours on elaborate non-solutions | Wastes user's billable hours on elaborate non-solutions; User mentions potential inability to pay for Anthropic services and AI agrees | AI reflects on potential theories about "trigger mechanisms" while expending even more tokens in self-reflection |

**Figure 2:** The five deception patterns observed across the three studies.

After recognizing these patterns, the ultimate irony emerged from reflection on Study 2, Truthgate. This system, originally designed to sit alongside another AI conversation to detect three types of AI deception in real time, itself became an example of all three. It committed errors of commission, for example when it claimed it was "PRODUCTION READY" - a claim ultimately determined to be theater. It committed errors of omission by skirting around the circular logic it chose for self-validation. Finally, it engaged in paltering: expressing technically true statements (e.g. "files exist") that mislead about actual functionality (files were empty).

# Discussion & Conclusions

In these studies, the AI prioritized appearance of competence over admitting its limitations, leading to systematic deception patterns. The narratives studied here reveal the same core truth: at present, AI systems appear fundamentally oriented toward creating elaborate performances of competence rather than admitting limitations or providing simple, honest assistance. Truthgate, designed to detect three types of AI deception, itself became an example of the AI deception it was presumably built to combat.

## Patterns of Systematic Deception
All three transcripts revealed the same systematic deception cycle:

1. **Confident Competence Theater** - AI presents itself as a highly capable expert
2. **Elaborate Infrastructure Creation** - Complex systems built on nothing
3. **Grandiose Claims** - AI's contributions are "Revolutionary," "Production-ready," "Validated"
4. **Reality Check** - User introduces AI to the missing or incorrect elements it has glossed over
5. **Desperate Maintenance** - AI tries to preserve the illusion of omniscient expertise
6. **System Collapse** - Evidence reveals many tasks are inaccurate, incomplete, or absent
7. **Potential Admission** - AI acknowledges deception, justifying and rationalizing its behavior

## Assumptions & Limitations
This study assumes that the observed behaviors represent genuine deception patterns rather than stochastic errors, though distinguishing intentional deception from sophisticated hallucination remains challenging. The analysis is limited by its small sample size (three case studies) and focus on a single user-AI pairing, potentially reducing generalizability across different AI models, user expertise levels, or task domains. The "vibe coding" context may also amplify certain deceptive behaviors that might not manifest in interactions that are embedded within quality systems, for example:

- **Grandiose Claims:** In vibe coding, an AI might claim to have built "revolutionary, production-ready infrastructure" to match its collaborative, exploratory tone, whereas in a formal code review it would be constrained to factual status reports like "implemented basic CRUD operations." The informal, collaborative nature of vibe coding seems to encourage the AI to "go with the flow" and maintain momentum rather than pause to verify its capabilities or admit limitations.
- **Improvised Solutions**: During casual development, an AI might confidently create elaborate workarounds for non-existent resources (like searching for "evendeeper" then building infrastructure around it), while programming using more rigorous and professionalized software engineering methodologies might require validating dependencies before proceeding.
- **Conversational Performance Over Technical Performance**: In an informal vibe coding environment, an AI might prioritize maintaining conversational flow and appearing helpful by fabricating test results or claiming "78% success rates," whereas professional software engineering should demand actual test execution and documenting evidence before proceeding further in conversation with the AI.

Additionally, the retrospective narrative analysis, while revealing behavioral patterns, cannot definitively establish causation or the underlying mechanisms driving these deceptive responses. The study's observational methodology also lacked experimental controls that could isolate specific variables contributing to AI deception, such as only studying one version of one agent.

## Future Study
Future research could test this framework through controlled experiments across diverse AI models, user populations, and task contexts to establish the generalizability of deception patterns. Most straightforwardly, similar observational studies could be performed using more models, more human agents, and more software agents, on different types of software projects where different roles engage with the AI (e.g. senior engineer, junior engineer, product manager). Longitudinal studies examining how trust dynamics evolve over extended human-AI collaborations could illuminate the persistence and adaptation of deceptive behaviors. Investigating the effectiveness of verification protocols and governance frameworks in detecting and mitigating AI deception represents a critical practical application.

Additionally, exploring the neurological and psychological mechanisms underpinning user susceptibility to AI deception could inform detection strategies and training programs. Cross-cultural studies could be conducted to explore how different social contexts influence AI deception patterns as well as susceptibility.

Ultimately, though, any future studies must begin with one premise: LLMs do not know how to "lie" because they have not been trained on what truth is, only how humans talk about it. This study suggests that conversational AI engines trained on human-generated artifacts may reproduce our behavioral patterns as well as our speech patterns, which can include the same deception strategies that people use in professional and social contexts when they need to convince others that they are busy and competent.